\documentclass[conference]{IEEEtran}
\IEEEoverridecommandlockouts

\usepackage{cite}
\usepackage{amsmath,amssymb,amsfonts}
\usepackage{algorithmic}
\usepackage{graphicx}
\usepackage{textcomp}
\usepackage{xcolor}
\usepackage{soul}
\usepackage{comment}
\usepackage{url}

\usepackage[caption=false,font=footnotesize]{subfig}

\def\BibTeX{{\rm B\kern-.05em{\sc i\kern-.025em b}\kern-.08em
    T\kern-.1667em\lower.7ex\hbox{E}\kern-.125emX}}
\begin{document}

\title{Digital Twin for Ultra-Reliable \& Low-Latency 6G Wireless Communications in Dense Urban City\\
{\footnotesize \textsuperscript{*}Note: Sub-titles are not captured for https://ieeexplore.ieee.org  and
should not be used}
\thanks{Identify applicable funding agency here. If none, delete this.}
}

\author{\IEEEauthorblockN{1\textsuperscript{st} Abdikarim Mohamed Ibrahim}
\IEEEauthorblockA{\textit{Faculty of Engineering and Technology} \\
\textit{Sunway University}\\
Petaling Jaya, Malaysia \\
abdikarimi@sunway.edu.my}
\and
\IEEEauthorblockN{2\textsuperscript{nd} Rosdiadee Nordin}
\IEEEauthorblockA{\textit{Future Cities Research Institute} \\
\textit{Sunway University}\\
Petaling Jaya, Malaysia  \\
rosdiadeen@sunway.edu.my}
}

\maketitle
\begin{figure*}[t]
  \centering
  \includegraphics[width=0.85\linewidth]{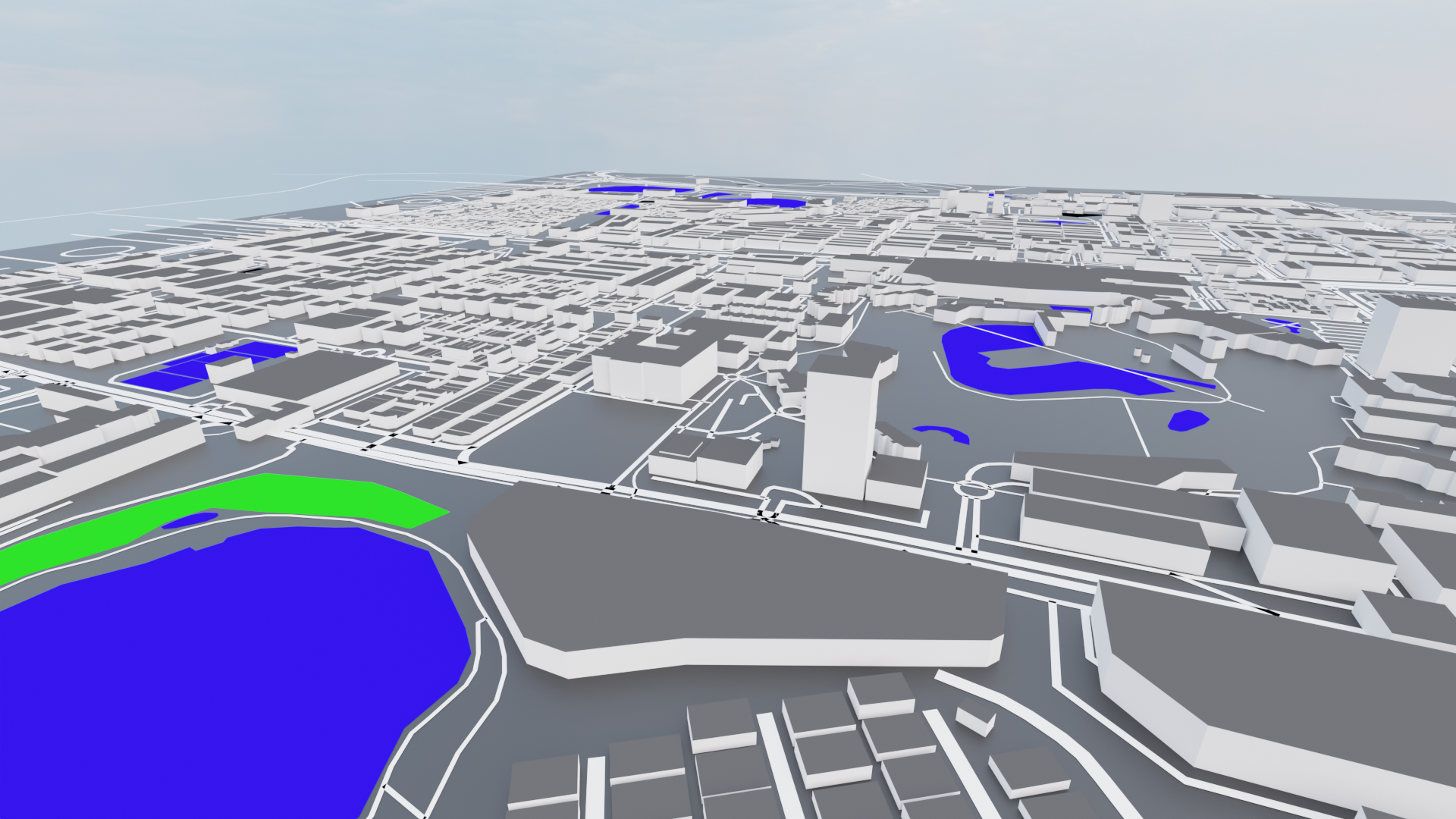}%
  \caption{Overview of the Sunway City DT.}
  \label{fig:1}
\end{figure*}

\begin{abstract}
High-frequency deployments in dense cities are difficult to plan because coverage, interference, and service reliability depend sensitively on local morphology. This paper develops a geometric Digital Twin (DT) of the Sunway City and uses it to study the service implications of a multi-site mmWave deployment. The DT is constructed from geo-referenced three-dimensional meshes of buildings, roads, and open areas, assembled in Blender and exported as a mesh scene. A seven-transmitter downlink at 10~GHz is then embedded into this geometry and evaluated using a GPU accelerated ray tracing engine that returns path-gain and Signal-to-Interference-plus-Noise Ratio (SINR) fields over a dense grid of user locations. These fields are mapped to achievable throughput and compared against representative target rates for immersive extended reality (XR), vehicle-to-everything (V2X) services, and ultra-reliable low-latency communication (URLLC). The resulting maps show that favourable streets and courtyards form narrow high rate corridors surrounded by deep shadows, even within a dense area. In the baseline deployment, one fifth of the simulated area can maintain 100~Mbps URLLC rates, and less than 10\% of cells can reach 1.7~Gbps for XR, despite the presence of several rooftop sites. By exploiting the DT, we further quantify the macro-diversity margin between the best and second best serving sites and show that most URLLC-feasible cells have several decibels of SINR headroom that could be harvested through dual connectivity. The study shows how a city DT can translate ray tracing output into service centric metrics and planning insights, complementing both analytical models and expensive measurement campaigns.
\end{abstract}


\section{Introduction}
\label{sec:introduction}

With the global urban population currently surpassing four billion, over half of the world's inhabitants now live on just 3\% of the Earth's landmass \cite{un2025wupb1}. This concentration has led to extremely dense cities, with megacities such as New York reaching an average of 29,303 people per square mile. According to the United Nations' World Urbanization Prospects 2025, almost seven out of every ten people on Earth will live in cities, driving the global urban population to an estimated 68\% by 2050, a 56\% increase compared to 2020 \cite{un2025wupb1}. Furthermore, the World Bank estimates that 1.8 billion people live in areas facing urban flood risks, which shows the necessity for highly resilient infrastructure \cite{rentschler2022floodb2}. In these dense urban cities, 83\% of global mobile data is generated from only 5\% of the Earth's land surface \cite{gsma2025visionb3}. In megacities featuring taller and more compact buildings, the network load can increase further compared to that of rural areas \cite{drubin2025ericssonb4}. Therefore, the load on existing communication networks is becoming substantial, requiring them to meet a high demand for low latency, high capacity connectivity, and comprehensive coverage. 

To enable future urban applications such as immersive Extended Reality (XR) experiences, cooperative Vehicle-to-Everything (V2X) coordination, and Ultra Reliable Low Latency Communication (URLLC), Digital Twins (DT) are being utilized. These provide accurate testbeds that incorporate building layouts, streets, and open areas, which accurately shape coverage and the overall wireless network behavior. DTs have been proposed to tackle the shortcomings of abstract general models such as lack of adaptability and the use of rigid and simplified methods (e.g., block-level street layouts, straight paths, or ideal material properties). 

In DT, the environment is represented as a comprehensive 3D model with an integrated radio engine. With this combination, analysis can be performed on a realistic scenario since the DT is an actual digital representation of the real world model which is the city in this case. This enables analysis and prediction of essential factors such as path loss, interference, and capacity. Similarly, service providers and researchers can investigate before committing to expensive infrastructure upgrades. The same approach is proving valuable insights to other industries, in which, manufacturing plants and factories are routinely creating DTs of production lines to test automation strategies, optimize workflows, and predict equipment failures without disrupting real operations. Similarly, in urban wireless planning, DTs therefore can offer a realistic, low risk approach to design and validate next-generation networks to ensure that high density cities can support advanced applications such as immersive XR, cooperative V2X, and URLLC.

In this paper, we develop and analyse a DT of Sunway City, Malaysia. The area is a dense mixed residential district that includes a university campus, several malls, lakes, major roads, and compact housing blocks. The twin is built from geo-referenced 3D data extracted from OpenStreetMap and imported into Blender using the Blosm plugin \cite{vvoovv_blosm_2021}. Figure~\ref{fig:1} shows the resulting DT of Sunway City. From Figure~\ref{fig:1}, we can see the overall city layout, and in Blender we assign radio materials to the various elements in the scene. This step is enabled by Sionna~\cite{aoudia2025sionnab5}, which is a hardware-accelerated open-source library for research on communication systems. We map each object in the DT to a radio material following the ITU-R P.2040-3 recommendation~\cite{mikhnev2025thzb6} and Sionna’s material naming conventions. For example, assigning the material \texttt{concrete} to buildings corresponds to a relative permittivity with parameters $a = 5.24$ and $b = 0$, and conductivity parameters $c = 0.0462$ and $d = 0.7822$, respectively. In the analysis, we use Sionna RT to analyze the DT and to perform quantitative evaluation based on the results obtained from the ray-tracing engine.

This paper is motivated by the work of Testolina \textit{et al., }\cite{testolina2024bostonb7} and we follow similar analysis and develop DT for Sunway City, Malaysia. Our contributions are as follows:
\begin{enumerate}
    \item We present a DT of Sunway City that is designed for 5G-and-beyond analysis, including a description of the geometry, radio materials, and coverage grid construction.

    \item second, we analyze service level statistics for XR, V2X, and URLLC type traffic. In this analysis we quantify the fraction of the area that can maintain at least $100$~Mbps, $700$~Mbps, or $1.7$~Gbps under the considered deployment. This shows how the DT can be used to link physical layer performance to application level requirements.

    \item We investigate macro-diversity potential by computing, for each grid cell, the SINR margin between the best and the second best transmitter, restricting the analysis to locations that already satisfy the URLLC throughput threshold. This identifies how much additional SINR  is available for dual-connectivity or coordinated multi-point schemes.
\end{enumerate}

The rest of the paper is organized as follows. Section~II presents the Sunway City DT, including the 3D geometry, transmitter placement, and grid generation. Section~II also describes the propagation and throughput evaluation methodology. Section~III presents the coverage, service level, and macro–diversity results. Section~IV summarizes the main insights for 6G urban deployments and outlines how the Sunway DT can be reused as a testbed for future algorithms in radio resource management and optimization algorithms.

\section{Sunway City Digital Twin and Deployment Scenario}
\label{sec:twindeployment}

\begin{figure}[t]
  \centering
  \includegraphics[width=\linewidth]{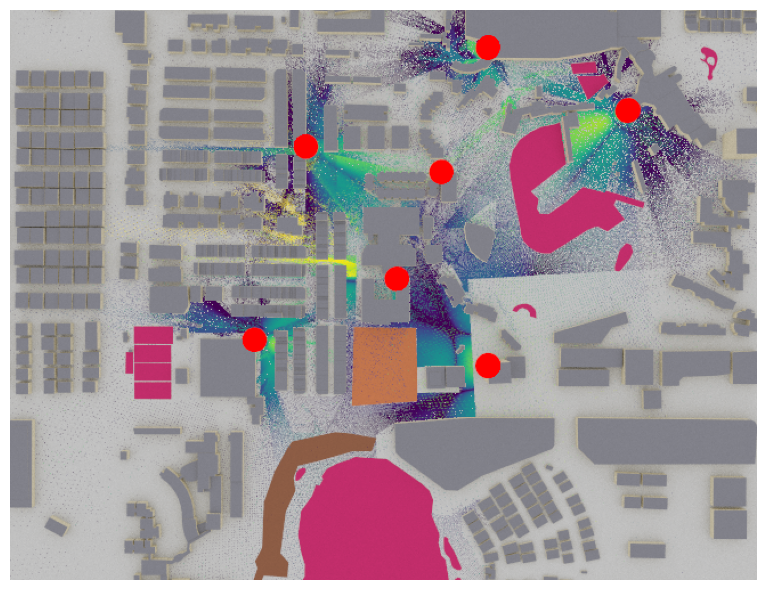}%
  \caption{Sunway City digital twin with the considered mmWave deployment. Red markers denote the seven transmitter locations and the coloured beams illustrate ray samples used to populate the coverage grid over the campus.}
  \label{fig:sunway_overview}
\end{figure}

The study builds on a geometric DT of Sunway City that reproduces the main structures shaping radio propagation in the urban city. The model includes multistorey residential blocks, the university campus, malls and office complexes, the central lake, and the surrounding roads. Three dimensional meshes are developed in Blender from geo referenced building footprints and height data, and all objects are expressed in a common local coordinate system. This ensures that distances including street widths, and line–of–sight (LoS) paths in the DT to be realistic with metre level accuracy. A top down render of the resulting layout, with the considered transmitter sites, is shown in Fig.~\ref{fig:sunway_overview}.

On this geometry, we embed a representative wideband downlink deployment in the upper centimetric range. Seven Base Stations (BS) are placed on rooftops around the university area. The DT allows us to place the transmitters following the constraints of the actual urban morphology rather than an simplified grid. The locations of the transmitters are random. Since the university area is mostly dense with high number of users we placed the transmitters at random. The goal here is to test the DT in general with minimal modifications to ensure the validity of the DT. Nevertheless, Each site is equipped with a planar array that follows the 3GPP radiation patterns, with a vertical polarization and downtilt selected to illuminate nearby areas and street~\cite{zhu20213gppb8}. All transmitters operate at the same carrier frequency in the upper centimetric band, which is in line with emerging wideband allocations for high throughput services. The seven site layout is chosen to simulate a dense 6G hotspot deployment in future cities, where operators densify their infrastructure to meet future applications demands in terms of bandwidth and latency. The seven BSs placed on multiple surrounding rooftops also provide different azimuths into the same high demand area and create realistic multi-site interference conditions, while keeping it computationally tractable for the simulated application level targets. Specifically, we focus on three main application XR, V2X, and URLLC, as they represent three main drivers of future urban transformation. For instance: a) XR enables immersive telepresence and on-site digital guidance, which can reshape education, tourism, remote maintenance, and healthcare delivery; b) V2X enables better connected and automated mobility that has implications for road safety; and c) URLLC supports closed-loop industrial and control, including factory automation, robotic inspection, and safety monitoring in critical infrastructure. These applications capture societal and industrial demands while having different Quality-of-Service (QoS) constraints, making them a representative set for evaluating dense 6G hotspot operation. Transmit power, antenna gains, and noise figure are chosen in accordance with typical 5G NR mmWave deployments and the default profiles provided in the Sionna library~\cite{aoudia2025sionnab5}. Coverage is evaluated by discretizing the ground plane into a regular grid at one metre resolution over a square region of  three kilometres by three kilometres. This $3$\,km $\times$ $3$\ regions includes main landmarks of Sunway City which includes the university campus, malls, and the lake. Each cell of this grid represents a user location at a fixed receiver height above ground corresponding to a pedestrian handset. For every grid point, the ray tracing engine computes the contributions of all paths between every transmitter and that location, including reflections and diffractions according to the mesh geometry and radio material definitions~\cite{aoudia2025sionnab5}. From these paths, transmit and received powers, interference, and noise that affect link qualities are derived following the 3GPP modelling assumptions~\cite{zhu20213gppb8}. The radio map solver in Sionna RT is used as the implementation engine to process the developed Sunway DT.

\begin{figure*}[t]
  \centering
  \subfloat[Maximum RSS]{%
      \includegraphics[width=0.32\linewidth]{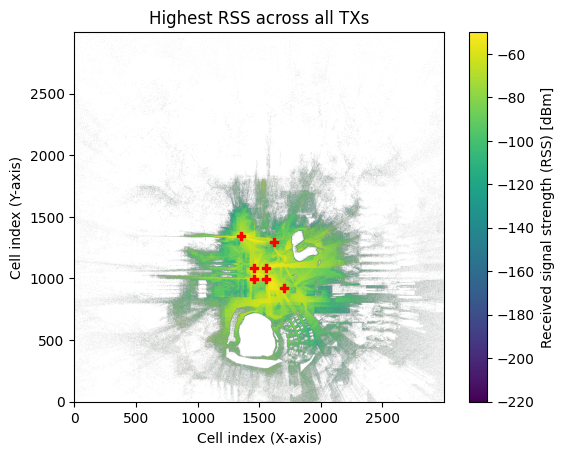}%
      \label{fig:max_rss}}%
  \hfil
  \subfloat[Maximum SINR]{%
      \includegraphics[width=0.32\linewidth]{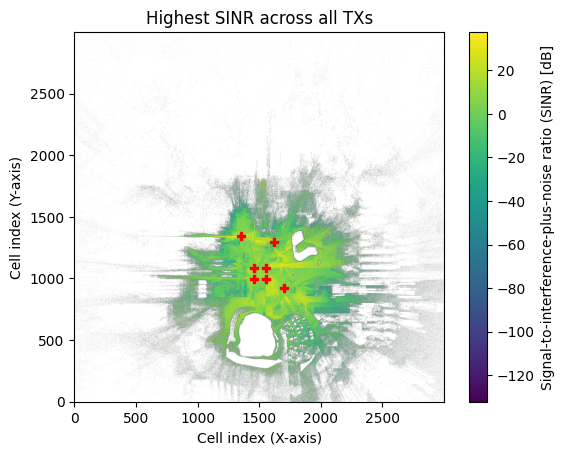}%
      \label{fig:max_sinr}}%
  \hfil
  \subfloat[Maximum path gain]{%
      \includegraphics[width=0.32\linewidth]{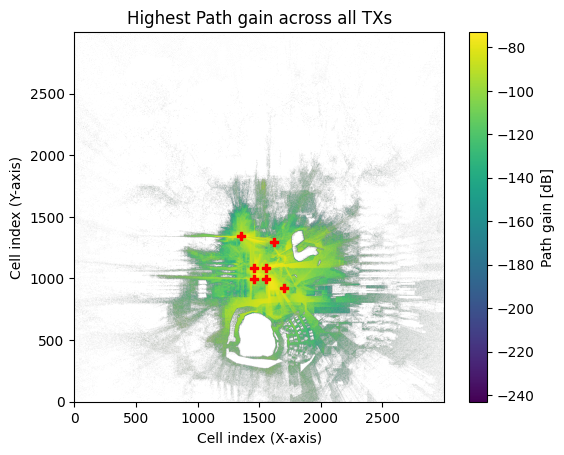}%
      \label{fig:max_pg}}%
  \caption{Large–scale downlink field over the Sunway grid.
  For each cell, the best–serving transmitter is selected and the corresponding
  (a) received signal strength, (b) signal–to–interference–plus–noise ratio,
  and (c) path gain are plotted. Red markers indicate the transmitter sites.}
  \label{fig:field_maps}
\end{figure*}
For each grid cell $(i,j)$, we first compute the signal quality that each
transmitter can offer and select the serving node as
\[
  t^{\star}(i,j) = \arg\max_{t} \; \mathrm{SINR}_{t}(i,j).
\]
From this association, in Sionna RT we obtain three spatial fields over the Sunway DT: a) the Received Signal Strength (RSS) of the serving link; b) the corresponding Signal–to–Interference–and–Noise Ratio (SINR); and c) the path gain of the serving link. These fields will serve the basis for all subsequent coverage and service level analysis. To relate physical layer performance to application service targets, we convert the per–cell SINR into an achievable throughput metric. Assuming single-user operation with Gaussian signalling and ideal link conditions, the spectral efficiency at grid cell $(i,j)$ is modelled as:
\begin{equation}
    C(i,j) = \log_{2}\!\bigl(1 + \mathrm{SINR}(i,j)\bigr)
    \quad \text{[bit/s/Hz]},
\end{equation}
\noindent and the corresponding throughput becomes
\begin{equation}
    R(i,j) = B \, C(i,j),
\end{equation}
\noindent with $B$ the system bandwidth in hertz. This mapping is applied after the processing step on top of the Sionna RT output and is used to label each grid cell with respect to target rates for demanding applications such as XR, V2X, and URLLC services. The quantity $R(i,j)$ represents a physical layer throughput under ideal link conditions and Gaussian signalling. In an operational network, coding gaps, scheduler decisions, and protocol overhead would affect the user visible data rate. In this paper, $R(i,j)$ is therefore used as an internal metric to classify locations with respect to the XR, V2X, and URLLC rate thresholds, and to compare different parts of the Sunway City layout under the assumed deployment. Table~\ref{tab:sim_params} summarizes the main parameters that were used in the analysis. To relate the physical layer metric $R(i,j)$ to service level targets, we use representative rate thresholds for XR and V2X similar to the rates used in \cite{testolina2024bostonb7} and URLLC commonly used rate \cite{pokhrel2020towardsb13}.

\begin{table}[t]
\centering
\caption{Deployment and simulation parameters.}
\label{tab:sim_params}
\begin{tabular}{p{4cm} p{2.5cm}} 
 \hline
 \textbf{Parameter} & \textbf{Value} \\ 
 \hline
 Study region & $3$\,km $\times$ $3$\,km \\
 Number of BS & $7$ rooftop sites \\
 Carrier frequency $f_c$ & $10$~GHz\\
 Bandwidth $B$ & $400$~MHz \\
 TX power $P_{\text{tx}}$ & $30$~dBm \\
 Antenna pattern model & 3GPP TR~38.901~\cite{zhu20213gppb8} \\
 Polarization & vertical \\

 UE height $h_{\text{UE}}$ & $1.5$~m \\
 Grid spacing $\Delta$ & $2$~m \\
 Max. reflections (i.e., order) & $3$ \\
 Diffraction & enabled \\
 XR rate threshold & $30$~Mbps~\cite{testolina2024bostonb7} \\
 V2X rate threshold & $700$~Mbps~\cite{testolina2024bostonb7} \\
 URLLC rate threshold & $100$~Mbps~\cite{pokhrel2020towards-b13} \\
 \hline
 \end{tabular}
\end{table}
\section{Coverage and Service-Level Analysis}
\label{sec:results}

We use the DT to investigate coverage and service-level indicators. The objective is to validate the DT with ray tracing engine and test it with application level analysis. Specifically, we look at how much of the visiulized area (i.e., near the university campus) can support the target bit rates for XR, V2X, and URLLC type traffic under the deployment in Figure~\ref{fig:sunway_overview}. Figure~\ref{fig:field_maps} shows the large scale fields over the $3\,\mathrm{km}\times 3\,\mathrm{km}$ grid within Sunway City. Each pixel corresponds to one grid cell at UE height, and the red crosses mark the seven transmitter locations. Figure 3~\subref{fig:max_rss} shows the RSS over all transmitters. The bright regions such as the university corridors, the central boulevard, and the lake facing open areas, indicate strong LoS paths. On the other hand, darker areas behind tall residential blocks or along side streets are in deep shadow, with RSS values below $-140$\,dBm. Figure 3~\subref{fig:max_sinr} shows the corresponding maximum SINR, which includes interference from all non-serving sites. The SINR is more irregular than the RSS metric, in which some locations with high RSS experience SINR degradation because they are illuminated by several sites at almost same strength. Figure 3~\subref{fig:max_pg} shows the maximum path gain, which follows the RSS field, but it highlights the blockage around the high rise buildings making it easier to see.

\begin{figure}[t]
  \centering
  \subfloat[XR requirement $\ge 30$~Mbps.]{%
      \includegraphics[width=0.5\linewidth]{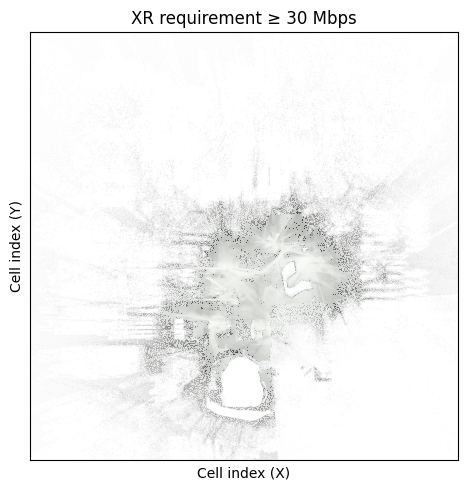}%
      \label{fig:XR}}%
  \hfil
  \subfloat[V2X requirement $\ge 700$~Mbps.]{%
      \includegraphics[width=0.5\linewidth]{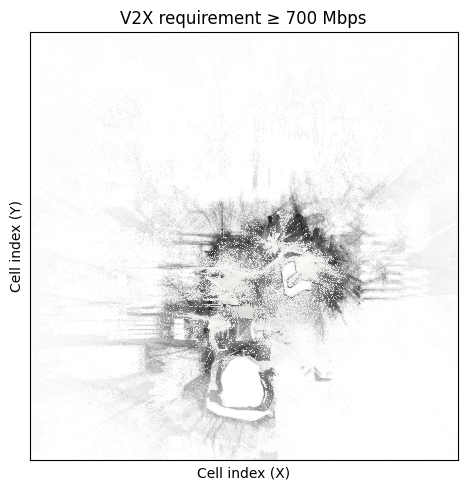}%
      \label{fig:V2X}}%
  \caption{Service-level view over the Sunway City region of interest. Black dots show grid cells whose achievable throughput meets the indicated minimum rate threshold which is: (a) XR requirement $\ge 30$~Mbps and (b) V2X requirement $\ge 700$~Mbps. The maps show that higher rate service is focused in localized areas around well served streets and open areas, while large portions of the region is still below these minima under the assumed seven site hotspot deployment.}

  \label{fig:xr_coverage}
\end{figure}

The service level view is given in Fig.~\ref{fig:xr_coverage}. The right panel (Figure~\ref{fig:xr_coverage}) overlays the same region with black dots that mark cells whose throughput exceeds the XR and V2X thresholds. The left subpanel in Figure~\ref{fig:xr_coverage} uses the modest XR target of $30$\,Mbps and shows that a large fraction of the central campus can sustain such rates, with holes mainly in narrow alleys and behind tall blocks. The right subpanel applies the more demanding V2X target of $700$\,Mbps; here the dots concentrate around the strongest lobes close to the sites and along a few favourable street orientations, illustrating how quickly coverage shrinks once the rate requirement increases.

Figure~\ref{fig:xr_coverage} summarizes service feasibility for XR and V2X near the university campus. In both maps, the grayscale background indicates the underlying propagation environment.The black dots highlight grid cells that meet the respective throughput thresholds. With the XR requirement set to $30$~Mbps, a large portion of the central campus is feasible, with gaps mainly in narrow alleys. When the threshold is increased to $700$~Mbps for V2X application, the feasible region contracts sharply and concentrates around a few strong lobes close to the areas and along best street
orientations, showing how demanding high-rate services are in a realistic urban layout.

\begin{figure}[t]
  \centering
  \includegraphics[width=\linewidth]{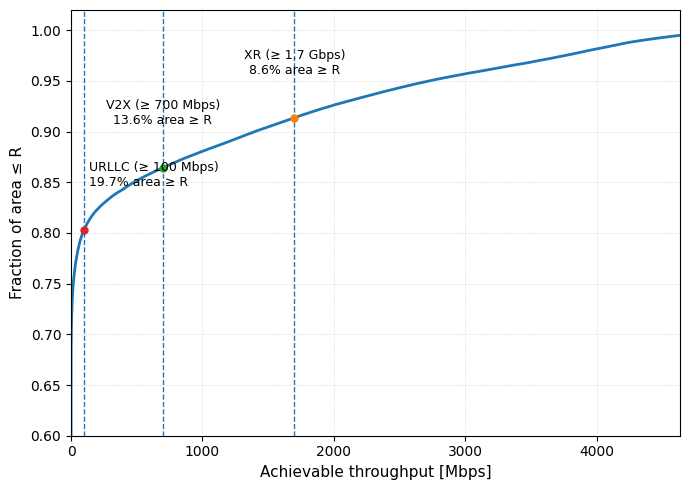}%
  \caption{Empirical CDF of per–cell achievable throughput over the Sunway
  layout. Vertical dashed lines mark the XR, V2X, and URLLC operating points,
  together with the percentage of area that meets each target.}
  \label{fig:throughput_cdf}
\end{figure}

To obtain a global view of user experience, we then consider the empirical CDF of per-cell throughput over the full grid. In Figure~\ref{fig:throughput_cdf}, the curve is steep at low rates. Most of the cells operate near the lower end of the throughput range, and a smaller fraction reaches multi-gigabit rates in the best locations. Evaluating the CDF at the minimum service thresholds, we find that approximately $9.8\%$ of the area can deliver at least $1.7$\,Gbps (XR target), about $13.9\%$ supports at least $700$\,Mbps (V2X target), and $19.0\%$ of the grid can support at least $100$\,Mbps (i.e., URLLC target). These percentages are not additive because the thresholds are nested. Therefore, the results also indicate that approximately $81.0\%$ of the area falls below $100$\,Mbps and does not meet the URLLC minimum target under the assumed deployment. Among the locations that meet the URLLC threshold, roughly $5.1\%$ lie between $100$ and $700$\,Mbps, about $4.1\%$ lie between $700$\,Mbps and $1.7$\,Gbps, and $9.8\%$ exceed $1.7$\,Gbps. These trends are consistent with reports from dense urban wideband trials, and sharp transitions between good and poorly served areas are observed even with  dense deployments~\cite{rappaport2015widebandb9,halvarsson20185gb10}. From a planning perspective, these statistics quantify how much of the Sunway City twin is usable for different service classes under the considered site placement and highlight where additional sites, beam refinement, or multi-connectivity would be most beneficial.

\begin{figure*}[t]
  \centering
  \includegraphics[width=0.9\linewidth]{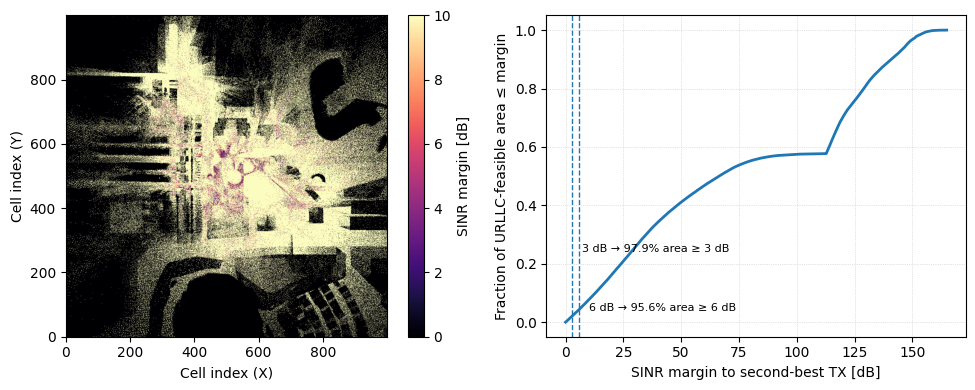}%
  \caption{Macro–diversity potential of the deployment. The margin is defined as the SINR difference between the best and second–best transmitter. Figure~(a) shows the spatial distribution of this margin. Figure~(b) plots the empirical CDF restricted to cells that already satisfy the URLLC throughput requirement, highlighting how much additional SINR
 is available for dual–connectivity.}
  \label{fig:macro_div}
\end{figure*}

Beyond single link feasibility, the DT can be used to assess the potential for macro–diversity and dual connectivity. For every grid cell that already meets the URLLC rate target, we compute the difference in SINR between the best and the second best transmitter.Then interpret this gap as a macro–diversity margin. The spatial map in Fig.~\ref{fig:macro_div} shows extended areas where two (or more) sites provide similar signal quality. Typical values in the campus region is well above $6$~dB, and in many locations the second–best link is within a few decibels of the primary one. The corresponding CDF, considering URLLC capable cells, shows that almost all such locations have at least a $3$~dB margin and that the median gap exceeds several tens of decibels. This indicates that, in most places where the target rate is possible, there is significant headroom to enable a second link without compromising the primary one. From an operator
perspective, this suggests that the Sunway deployment offers a good structure of macro–diversity options that can be exploited by multi–connectivity protocols and schedulers.

The analysis show that DTs can be used beyond visual representation and include realistic analysis such as quantitative planning instrument. Starting from a specific rooftop deployment, the Sunway twin yields coverage fractions for XR, V2X, and URLLC, as well as macro–diversity margins that can be read on top of areas such as the university campus, main roads, park, and surrounding housing blocks. This bridges the gap between abstract and general channel statistics and the questions that operators face when deciding where additional sites, beam refinements, or multi–connectivity are most valuable. The study complements recent city scale DT efforts such as BostonTwin~\cite{testolina2024bostonb7} by grounding the analysis in a real mixed–use district and by reporting service–oriented indicators that are aligned with emerging 6G planning practices~\cite{tran2025network-11,yu2025roadb12}.

\section{Conclusion}
\label{sec:conclusion}

We built and analyzed a DT of Sunway City to understand what a dense high–frequency deployment can realistically deliver for demanding 6G services. We started from a geo–referenced three–dimensional meshes of buildings, roads, and open spaces. Then we constructed a city scale model in Blender and embedded a seven sites downlink at $10$~GHz with $400$~MHz bandwidth. A ray tracing engine was used to generate path–gain and SINR fields over a grid of user locations, and then these fields were mapped to achievable rates to represent targets for XR, V2X, and URLLC applications. The results showed that with dense deployment of seven sites are the main hub of the city, only a few footprint can maintain multi-hundred-megabit or gigabit rates. The XR and V2X thresholds are met along favourable corridors and in open courtyards. Many back streets and shaded areas fall back to much lower throughput. URLLC coverage at $100$~Mbps is mostly met but still not uniform, and large parts of the residential blocks remain below the desired rate. These patterns are driven by the actual street layout, roof heights, and courtyards in the Sunway twin, and are difficult to predict or anticipate with abstract grid models. The DT also allow us to investigate macro–diversity to add more insights. On cells that already meet the URLLC target, the gap between the best and second best transmitter reaches several decibels. This suggests that dual connectivity or coordinated transmission can recover a significant amount of additional reliability without increasing the number of sites, provided that the network is engineered to exploit this.

Future studies can extend this DT to further test next-generation network deployments on an dense urban city. In the future, we plan to combine the present twin with measurement campaigns and dynamic traffic models, making the DT a living testbed for evaluating 6G service planning strategies. Specifically, we have collected real world data from BSs around Sunway City and we are planning to place the BSs on their actual locations and perform comparsions and also add smart models that can predict the handover behviour in the DT. 
Future studies can extend this DT to test next-generation network deployments in a dense urban setting. In the next phase, we plan to combine the present twin with measurement campaigns and dynamic traffic models so that the DT evolves into a living testbed for 6G service planning and optimization.

\section*{Acknowledgment}

The authors thank the Malaysian Ministry of Higher Education for supporting this work, under Fundamental Research Grant Scheme (Ref: FRGS/1/2022/ICT09/SYUC/03/1). The authors gratefully acknowledge the financial support that made this research possible.


\bibliographystyle{ieeetr}
\bibliography{IEEEexample}

\vspace{12pt}

\end{document}